\documentclass{jfm}
\usepackage{graphicx}
\usepackage{epstopdf, epsfig,subfloat,subfig}

\usepackage{amsfonts}
\usepackage{amsmath}
\usepackage{amssymb,color}

\shorttitle{A comment on `macroscale model for gas slip flow in porous media'}
\shortauthor{Wu, Ho, Germanou, Gu, Liu, Xu, and Zhang } 

\title{A comment on `An improved macroscale model for gas slip flow in porous media'}

\author{Lei Wu\aff{1}
	\corresp{\email{lei.wu.100@strath.ac.uk}},
	Minh Tuan Ho\aff{1},
	Lefki Germanou\aff{1},
	Xiao-Jun Gu\aff{2},
	Chang Liu\aff{3},
	Kun Xu\aff{3},
	\and Yonghao Zhang\aff{1}
	}

\affiliation{
	\aff{1}James Weir Fluids Laboratory, Department of Mechanical and Aerospace Engineering, University of Strathclyde, Glasgow G1 1XJ, UK
	\aff{2}Computational Science and Engineering Department, STFC Daresbury Laboratory, Warrington WA4 4AD, UK
	\aff{3}Department of Mathematics, Hong Kong University of Science and Technology, Clear Water Bay, Hong Kong, China
}

\begin{document}

\maketitle

\begin{abstract}

In a recent paper by Lasseux, Vald\'{e}s-Parada and Porter (J.~Fluid~Mech. \textbf{805} (2016) 118-146), it is found that the apparent gas permeability of the porous media is a nonlinear function of the Knudsen number. However, this result is highly questionable, because the adopted Navier-Stokes equations and the first-order velocity-slip boundary condition are first-order (in terms of the Knudsen number) approximations of the Boltzmann equation and the kinetic boundary condition for rarefied gas flows. Our numerical simulations based on the Bhatnagar-Gross-Krook kinetic equation and regularized 20-moment equations prove that the Navier-Stokes equations with the first-order velocity-slip boundary condition are only accurate at a very small Knudsen number limit, where the apparent gas permeability is a linear function of the Knudsen number.
\end{abstract}


\section{Introduction}

The permeability of porous media is an important parameter to predict the unconventional gas production. For laminar flows in highly permeable porous media, the Darcy's law states that the volume flow rate $Q$ is proportional to the pressure gradient $\nabla{p}$:
\begin{equation}\label{Darcy}
Q=-k_\infty\frac{A}{\mu}\nabla{p},
\end{equation}
where $A$ is the cross-section area of the flow, $\mu$ is the shear viscosity of the fluid, and $k_\infty$ is the permeability of a porous medium that is independent of the fluid. For this reason, $k_\infty$ is known as the intrinsic permeability.

For gas flows in low permeable porous media, however, the measured permeability is larger than the intrinsic permeability and increases with the reciprocal mean gas pressure $\bar{p}$~\citep{Klinkenberg1941}. In order to distinguish it from the intrinsic permeability, the permeability is called the apparent  permeability $k_a$, which can be expressed as:
\begin{equation}\label{Klinkenberg}
{k_a}={k_\infty}\left(1+\frac{b}{\bar{p}}\right),
\end{equation} 
where $b$ is  the correction factor.

The variation of the apparent permeability with respect to the mean gas pressure is due to the rarefaction effects, where infrequent collisions between gas molecules not only cause the gas slippage at the solid surface, but also modify the constitution relation between the stress and strain-rate~\citep{henning}. The extent of rarefaction is characterized by the Knudsen number $Kn$ (i.e. the ratio of the mean free path $\lambda$ of gas molecules to the characteristic flow length $L$):
\begin{equation}
Kn=\frac{\lambda}{L}, \\~\\ \textrm{and} \\~\\  \lambda=\frac{\mu(T_0)}{\bar{p}}\sqrt{\frac{\pi{RT_0}}{2}},
\end{equation}
where $\mu(T_0)$ is the shear viscosity of the gas at a reference temperature $T_0$, and $R$ is the gas constant. Gas flows can be classified into four regimes\footnote{Note that this partition of flow regime is roughly true for the gas flow between two parallel plates with a distance $L$; for gas flows in porous media, the region of $Kn$ for different flow regimes may change.}: continuum flow $(Kn\lesssim0.001)$ in which Navier-Stokes equations (NSEs) can be used; slip flow $(0.001<Kn\lesssim0.1)$ where NSEs with appropriate velocity-slip/temperature-jump boundary conditions may be used; transition flow $(0.1<Kn\lesssim10)$ and free-molecular flow $(Kn>10)$, where NSEs break down and the Boltzmann equation is used to describe rarefied gas flows~\citep{CE}.

Recently, based on NSEs with the first-order velocity-slip boundary condition (FVBC), \cite{Didier2016} found that the apparent gas permeability of the porous media is a nonlinear function of $Kn$. This result, however, is questionable, because the NSEs were used  beyond its validity. Through our theoretical analysis and numerical calculations, we show that NSEs with FVBC can only predict the apparent permeability of porous media to the first-order accuracy of $Kn$.

\section{State of the problem}\label{Sec2}

\begin{figure}
	\centering
{	\includegraphics[scale=0.55,viewport=140 270 390 525,clip=true]{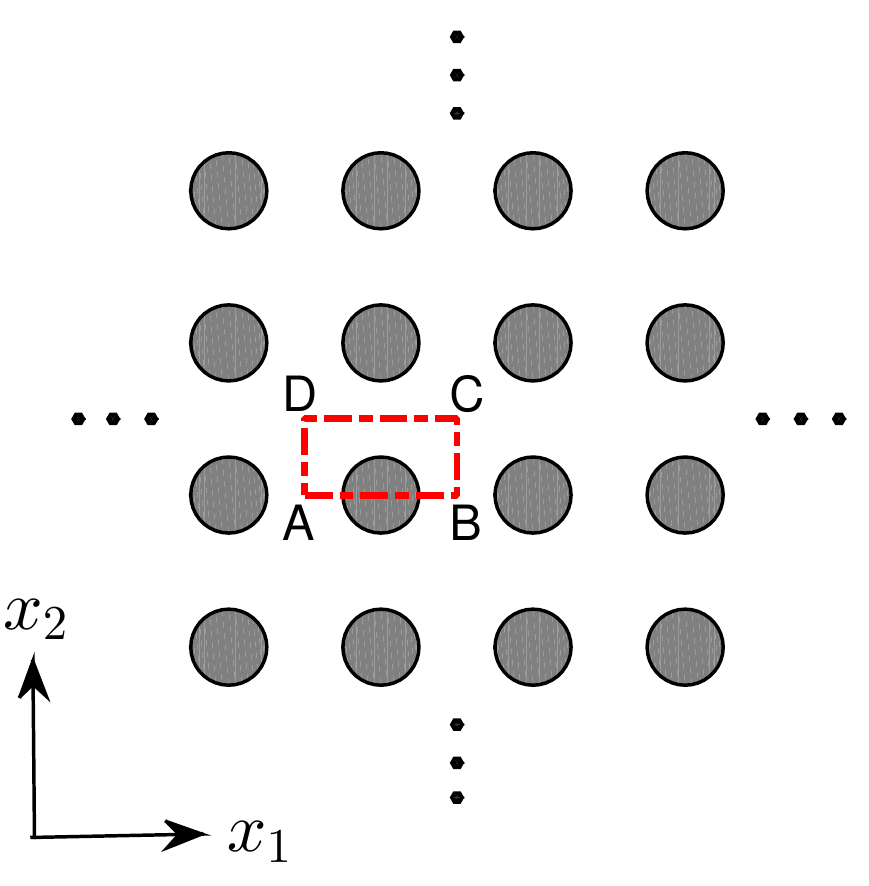}}

	\caption{A two-dimensional porous medium consisting of a periodic array of discs. A, B, C, and D are the four corners of the unit rectangular cell (computational domain). }
	\label{GEO}
\end{figure}

Consider a gas flowing through the periodic porous media. Suppose the geometry along the $x_3$-direction is uniform and infinite, the gas flow is effectively two-dimensional and can be studied in a unit rectangular cell ABCD, with appropriate governing equations and boundary conditions; one example of the porous medium consisting of a periodic array of discs is shown in Fig.~\ref{GEO}. We are interested in how the apparent permeability varies with the Knudsen number.

\subsection{The mesoscopic description: the gas kinetic theory}

The Boltzmann equation is fundamental in the study of rarefied gas dynamics from the continuum to the free-molecular flow regimes, which uses the  distribution function $f(t,\textbf{x},\textbf{v})$ to describe the system state:
\begin{equation}\label{Boltzmann}
\frac{\partial f}{\partial t}+v_1\frac{\partial f}{\partial x_1}+v_2\frac{\partial f}{\partial x_2}+v_3\frac{\partial f}{\partial x_3}=\mathcal{C}(f),
\end{equation}
where $\textbf{v}=(v_1,v_2,v_3)$ is the three-dimensional molecular velocity normalized by the most probable speed $v_m=\sqrt{2RT_0}$, $\textbf{x}=(x_1,x_2,x_3)$ is the spatial coordinate normalized by the length $L$ of the side AB, $t$ is the time normalized by $L/v_m$, $f$ is normalized by $\bar{p}m/v_m^3RT_0$,  while $\mathcal{C}$ is the Boltzmann collision operator. 

In order to save the computational cost, $\mathcal{C}$ is usually replaced by the relaxation-time approximation~\citep{Bhatnagar1954}, resulting in the Bhatnagar-Gross-Krook (BGK) equation. Numerical simulation for the Poiseuille flow between two parallel plates shows that the BGK equation can yield  accurate mass flow rates when the gas flow is not in the free-molecular regime~\citep{Sharipov2009}.

When the porous medium is so long that the pressure gradient is small, the BGK equation can be linearized. The distribution function is expressed as $f=f_{eq}(1+h)$, where $f_{eq}={\pi^{-{3}/{2}}}{\exp(-|\textbf{v}|^2)}$ is the equilibrium distribution function, and the perturbation $h$ is governed by~\citep{GraurVacuum2012}:
\begin{equation}\label{BGK}
\frac{\partial h}{\partial t}+v_1\frac{\partial h}{\partial x_1}+v_2\frac{\partial h}{\partial x_2}=\frac{\sqrt{\pi}}{2Kn}\left[\varrho+2u_1v_1+2u_2v_2+\tau\left(|\textbf{v}|^2-\frac{3}{2}\right)-h\right],
\end{equation}
where macroscopic quantities such as the perturbed density $\varrho$, the velocity $u_1$ and $u_2$, and the perturbed temperature $\tau$ are calculated as
\begin{equation}
\begin{split}
\varrho=\int{h}f_{eq}d\textbf{v},\ \ \ \ 
(u_1,u_2)=\int(v_1,v_2){h}f_{eq}d\textbf{v},\ \ \ \ 
\tau=\frac{2}{3}\int |\textbf{v}|^2{h}f_{eq}d\textbf{v}-\varrho,
\end{split}
\end{equation}

The kinetic equation~\eqref{BGK} has to be supplied with the boundary condition. Suppose the pressure gradient is along the $x_1$ direction, on the inlet and outlet of the computational domain ABCD (the coordinates of the four corners A, B, C, and D are $(-0.5,0), (0.5,0), (0.5, 0.5)$, and $(-0.5,0.5)$, respectively), the pressure gradient is applied and the periodic condition for the flow velocity is used~\citep{GraurVacuum2012}:
\begin{equation}
h\left(\mp0.5,x_2,v_1,v_2, v_3\right)=\pm1+h\left(\pm0.5,x_2,v_1,v_2,v_3\right), \text{ when } v_1\gtrless0,
\end{equation} 
at the lines AB and CD, the specular reflection boundary condition is used to account for the symmetry: $h\left(x_1,0,v_1,v_2, v_3\right)=h\left(x_1,0,v_1,-v_2, v_3\right)$ when $v_2>0$, and $h\left(x_1,0.5,v_1,v_2, v_3\right)=h\left(x_1,0.5,v_1,-v_2, v_3\right)$ when $v_2<0$, 
while at the solid surface, the diffuse boundary condition is used~\citep{Maxwell1879}:
\begin{equation}\label{diffuse}
h(x_1,x_2,\textbf{v})=\frac{2}{\pi} \int_{v_n'<0} |v_n'| h(x_1,x_2,\textbf{v}')\exp(-|\textbf{v}'|^2) d\textbf{v}',
\end{equation}
where $v_n$ is the normal velocity vector at the solid surface.



The apparent gas permeability, which is normalized by $L^2$, is calculated by
\begin{equation}
k_a=4\sqrt{\frac{1}{\pi}}Kn\int_0^{1/2}u_1(x_2)dx_2.
\end{equation}

\subsection{The macroscopic description: NSEs and moment equations}\label{macro_equation}

Historically, the state of a gas is first described by macroscopic quantities such as the density $\rho$, velocity $u_i$, and temperature $T$; and its dynamics is described by the Euler equations or NSEs (based on the empirical Newton's law for stress and the Fourier's law for heat flux). These equations, however, can be derived rigorously from the Boltzmann equation, at various order of approximations.

By taking the velocity moments of the Boltzmann equation~\eqref{Boltzmann}, the five macroscopic quantities are governed by the following equations: 
\begin{eqnarray}
\frac{\partial \rho}{\partial t}+\frac{\partial \rho {u_i}}{\partial {x_i}} = 0, \label{MASS} \\
\frac{\partial \rho u_i}{\partial t}+\frac{{\partial \rho {u_i}{u_j}}}{{\partial {x_j}}} + \frac{{\partial {\sigma _{ij}}}}{{\partial {x_j}}} =  - \frac{{\partial p}}{{\partial {x_i}}}, \label{MOMENTUM}\\
\frac{{\partial \rho T}}{{\partial t}} + \frac{{\partial \rho {u_i}T}}{{\partial {x_i}}} + \frac{2}{{3R}}\frac{{\partial {q_i}}}{{\partial {x_i}}} =  - \frac{2}{{3R}}\left( {p\frac{{\partial {u_i}}}{{\partial {x_i}}} + {\sigma _{ij}}\frac{{\partial {u_j}}}{{\partial {x_i}}}} \right). \label{ENERGY}
\end{eqnarray}

However, the above equations are not closed, since expressions for the shear stress $\sigma_{ij}$ and heat flux $q_i$ are not known. One way to close~\eqref{MASS}-\eqref{ENERGY} is to use the Chapman-Enskog expansion, where the distribution function is expressed in the power series of $Kn$~\citep{CE}:
\begin{equation}
f=f^{(0)}+Kn f^{(1)}+Kn^2 f^{(2)}+\cdots,
\end{equation}
where $f^{(0)}$ is the equilibrium Maxwellian distribution function. When $f=f^{(0)}$, we have $\sigma_{ij}= q_i=0$, and~\eqref{MASS}-\eqref{ENERGY} reduce to the Euler equations. When the distribution function is truncated at the first-order of $Kn$, that is, 
\begin{equation}\label{NS_fist}
f=f^{(0)}+Kn f^{(1)},
\end{equation}
we have 
\begin{equation}\label{GTMNSF}
\sigma_{ij} =-2\mu \frac{\partial u_{<i}}{\partial x_{j>}} \\~\\ \textrm{and} \\~\\  q_i = -\frac{15}{4}R\mu \frac{\partial T}{\partial x_i},
\end{equation}
and \eqref{MASS}-\eqref{ENERGY} reduce to NSEs. When $f=f^{(0)}+Kn f^{(1)}+Kn^2 f^{(2)}$, Burnett equations can be derived.

Alternatively, following the method of~\cite{Grad1949}, 13-, 20- and 26-moment equations~\citep{Struchtrup2003,Gu2009} can be derived from the Boltzmann equation to describe flows at different levels of rarefaction. Here the regularized 20-moment (R20) equations are used, which, in addition to \eqref{MASS}-\eqref{ENERGY}, include governing equations for the high-order moments $\sigma_{ij}$, $q_i$, and $m_{ijk}$:
\begin{equation}\label{STRESS}
\frac{\partial \sigma_{ij}}{\partial t}+\frac{{\partial {u_k}{\sigma _{ij}}}}{{\partial {x_k}}} + \frac{{\partial {m_{ijk}}}}{{\partial {x_k}}} = - \frac{p}{\mu }{\sigma _{ij}} - 2p\frac{{\partial {u_{ < i}}}}{{\partial {x_{j > }}}}  - \frac{4}{5}\frac{{\partial {q_{ < i}}}}{{\partial {x_{j > }}}} - 2{\sigma _{k < i}}\frac{{\partial {u_{j > }}}}{{\partial {x_k}}},
\end{equation}
\begin{eqnarray}\label{HFLUX}
\frac{\partial q_i}{\partial t} 
 &+& \frac{{\partial {u_j}{q_i}}}{{\partial {x_j}}} 
 + \frac{1}{2}\frac{{\partial {R_{ij}}}}{{\partial {x_j}}}
 = - \frac{2}{3}\frac{p}{\mu }{q_i}
- \frac{5}{2}p\frac{{\partial RT}}{{\partial {x_i}}} 
+ \frac{{{\sigma _{ij}}}}{\rho }\left( {\frac{{\partial p}}{{\partial {x_j}}} + \frac{{\partial {\sigma _{jk}}}}{{\partial {x_k}}}} \right)  
-  RT\frac{{\partial {\sigma _{ij}}}}{{\partial {x_j}}} \nonumber \\ 
&-& \frac{7}{2}{\sigma _{ij}}\frac{{\partial RT}}{{\partial {x_j}}} 
- \left( {\frac{2}{5}{q_i}\frac{{\partial {u_j}}}{{\partial {x_j}}} + \frac{7}{5}{q_j}\frac{{\partial {u_i}}}{{\partial {x_j}}} + \frac{2}{5}{q_j}\frac{{\partial {u_j}}}{{\partial {x_i}}}} \right) - {m_{ijk}}\frac{{\partial {u_j}}}{{\partial {x_k}}} - \frac{1}{6}\frac{{\partial \Delta }}{{\partial {x_i}}},
\end{eqnarray}
\begin{eqnarray}\label{MIJK}
\frac{\partial m_{ijk}}{\partial t} &+& \frac{{\partial {u_l}{m_{ijk}}}}{{\partial {x_l}}} + \frac{{\partial {\phi _{ijkl}}}}{{\partial {x_l}}} =
 - \frac{3}{2}\frac{p}{\mu }{m_{ijk}} - 3\frac{{\partial RT{\sigma _{ < ij}}}}{{\partial {x_{k > }}}}
 - \frac{{12}}{5}{q_{ < i}}\frac{{\partial {u_j}}}{{\partial {x_{k > }}}} \nonumber \\
 &+& 3\frac{{{\sigma _{ < ij}}}}{\rho }\left( {\frac{{\partial p}}{{\partial {x_{k > }}}} + \frac{{\partial {\sigma _{k > l}}}}{{\partial {x_l}}}} \right).
\end{eqnarray}

The constitutive relationships between the unknown higher-order moments ($R_{ij}$, $\Delta$ and $\phi_{ijkl}$) and the lower-order moments were given by Structrup \& Torrilhon (2003) and Gu \& Emerson (2009) to close~\eqref{MASS} to~\eqref{MIJK}. For linearized flows, it is adequate to use the linear gradient transport terms only and they are:
\begin{equation}
{\phi _{ijkl}} =  - \frac{{4\mu }}{{{C_1}\rho }}\frac{{\partial {m_{ < ijk}}}}{{\partial {x_{l > }}}}, \ \ \ \ 
{R_{ij}} =  - \frac{24}{5}\frac{\mu}{p} RT\frac{{\partial {q_{ < i}}}}{{\partial {x_{j > }}}}, \ \ \ \ 
\Delta  =  - 12\frac{\mu }{{p}} RT\frac{{\partial {q_k}}}{{\partial {x_k}}},
\end{equation}
where the collision constant $C_1$ is  2.097 for Maxwell molecules~\citep{Gu2009}.

Macroscopic wall boundary conditions were obtained from the diffuse boundary condition~\citep{Maxwell1879}. In a frame where the coordinates are attached to the wall, with $n_i$ the normal vector of the wall pointing towards the gas and $\tau_i$ the tangential vector of the wall, the velocity-slip parallel to the wall $u_{\tau}$ and temperature-jump conditions are:
\begin{eqnarray}
{u_\tau } =  - \sqrt {\frac{{\pi RT}}{2}} \frac{{{\sigma _{n\tau }}}}{{{p_\alpha }}} - \frac{{5{m_{nn\tau }} + 2{q_\tau }}}{{10{p_\alpha }}}, \label{ho_velocity}\\
RT - R{T_w} =  - \sqrt {\frac{{\pi RT}}{2}} \frac{{{q_n}}}{{2{p_\alpha }}}  - \frac{{RT{\sigma _{nn}}}}{{4{p_\alpha }}} + \frac{{u_\tau ^2}}{4} - \frac{{75{R_{nn}} + 28\Delta }}{{840{p_\alpha }}} + \frac{{{\phi _{nnnn}}}}{{24{p_\alpha }}},
\end{eqnarray}
where ${p_\alpha } = p + \sigma _{nn}/{2} - (30R_{nn} + 7\Delta)/ 840RT - \phi _{nnnn}/24RT$. The rest of wall boundary conditions for higher-order moments are listed in Appendix~\ref{Wall_hob}. Note that the velocity-slip boundary condition~\eqref{ho_velocity} is also of higher-order due to the appearance of the higher-order moment $m_{nn\tau}$.

Following the above introduction, it is clear that NSEs with FVBC are only accurate to the first-order of $Kn$; therefore, any apparent gas permeability showing the nonlinear dependence with $Kn$ is highly questionable. The R20 equations are accurate to the third-order of $Kn$~\citep{henning}, which should give the some apparent permeability of the porous media as NSEs when $Kn\rightarrow0$, and be more accurate than NSEs as $Kn$ increases. Numerical simulations are also performed to demonstrate this.

\section{Numerical results}

We first investigate the rarefied gas through a periodic array of discs with the diameter $D$, as shown in Fig.~\ref{GEO}. Using NSEs and FVBC, when the porosity $\epsilon=1-\pi{D}^2/4$ is large, the slip-corrected permeability can be obtained analytically~\citep{Chai2011}:
\begin{equation}\label{per_large_por}
k_a=\frac{1}{8\pi\left(1+2\xi{Kn}\sqrt{\frac{\pi}{\phi}}\right)}\bigg[-\ln\phi-\frac{3}{2}+2\phi-\frac{\phi^2}{2}
+2\xi{Kn}\sqrt{\frac{\pi}{\phi}}\left(-\ln\phi-\frac{1}{2}+\frac{\phi^2}{2}\right)\bigg],
\end{equation}
where $\phi=1-\epsilon$ is the solid fraction and $\xi=1.016\sqrt{{4}/{\pi}}=1.146$ when the diffuse boundary condition~\eqref{diffuse} is used~\citep{Hadjiconstantinou2003slip}.  The intrinsic permeability is obtained when $\epsilon=0$.

\begin{figure}
	\centering
	\subfloat[]{\includegraphics[scale=0.6,viewport=30 10 625 285,clip=true]{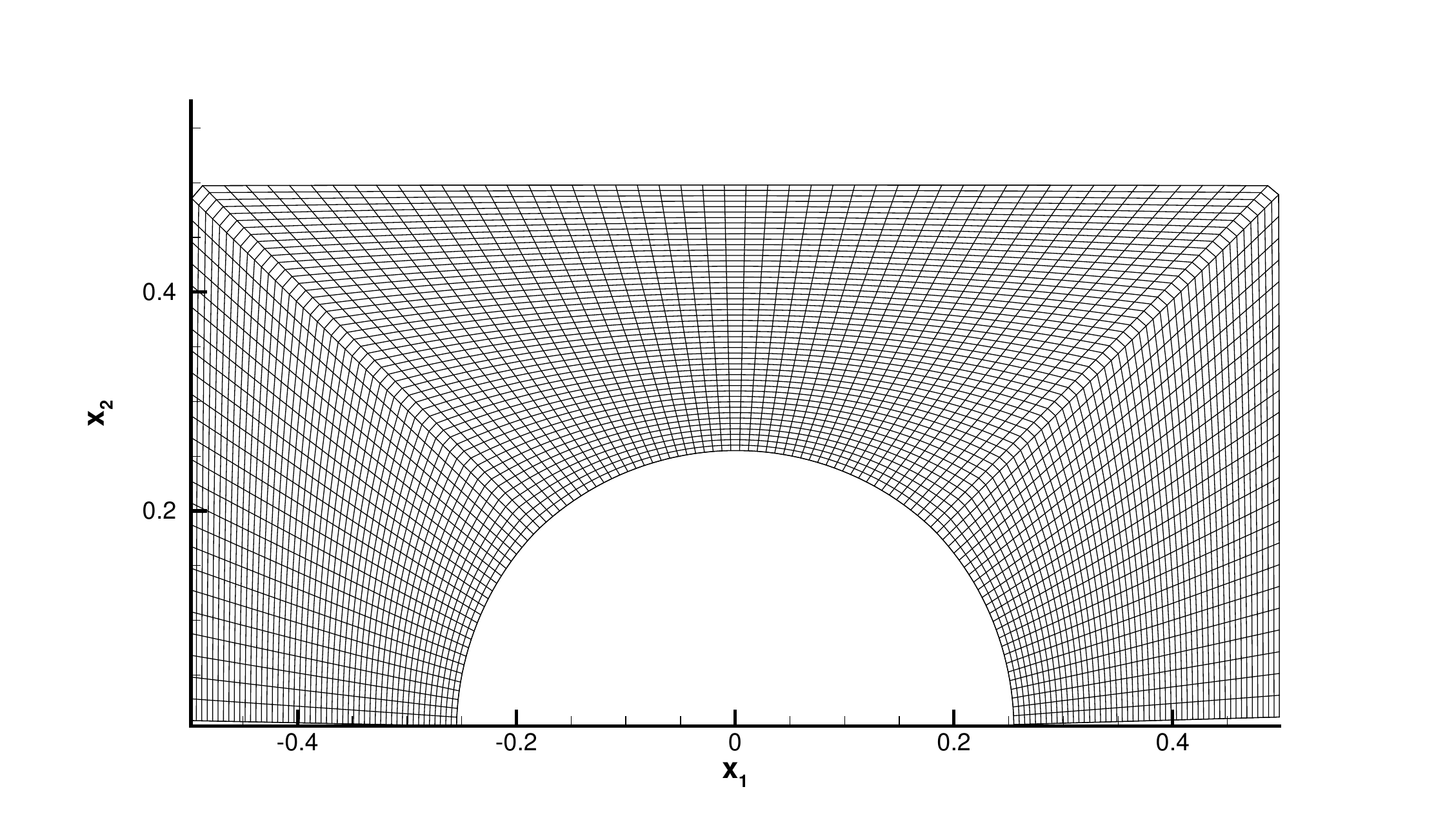}}\\
	
	\subfloat[]	{\includegraphics[scale=0.44,viewport=30 0 620 355,clip=true]{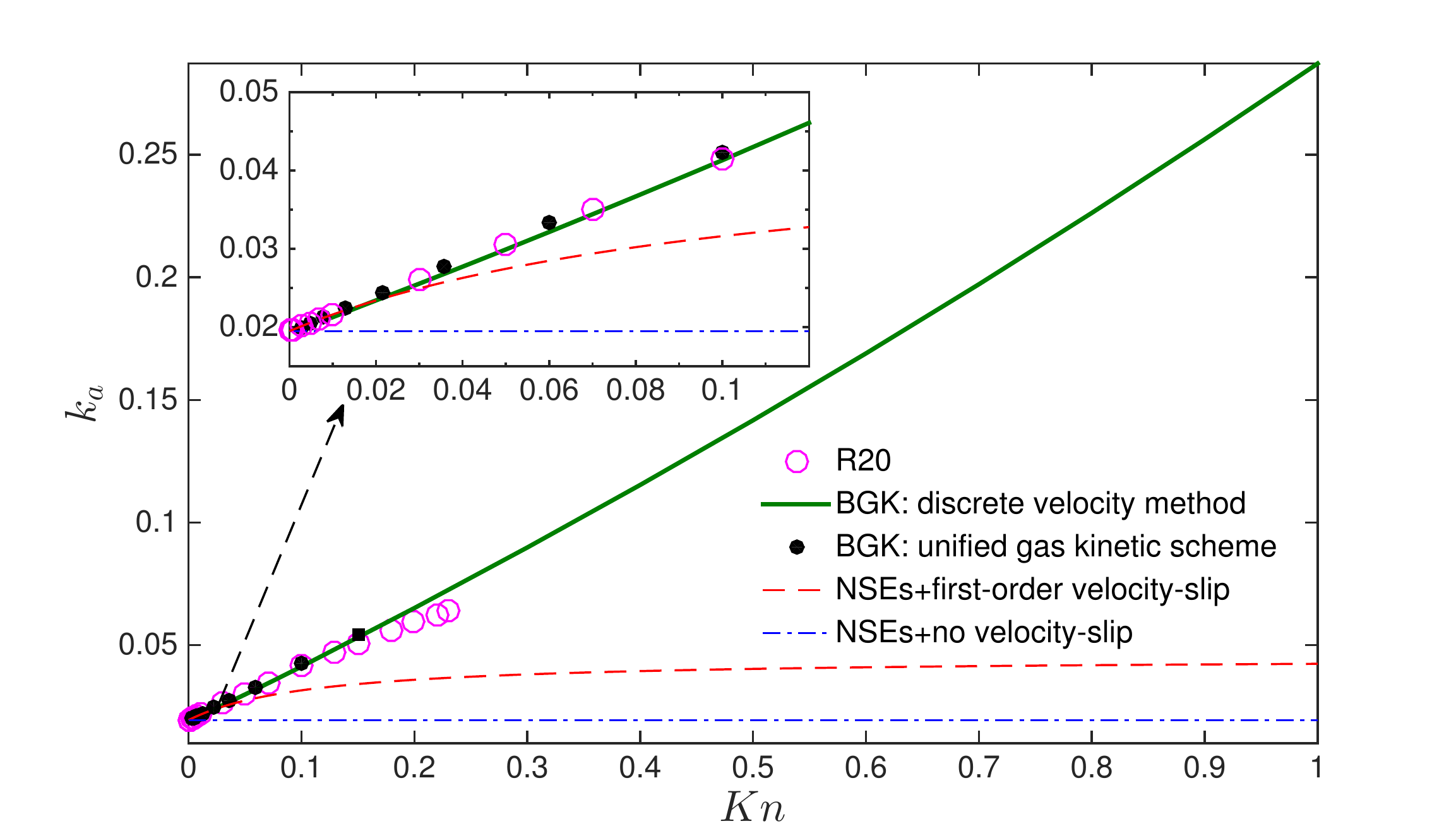}}
	
	\caption{(a) The body-fitted mesh used in the unified gas kinetic scheme, when the porosity of the porous medium in Fig.~\ref{GEO} is 0.8. For clarity, only $100\times50$ cells are shown. (b) The apparent permeability as a function of the Knudsen number. The solid line and dots are numerical results of the linearized BGK solved by the discrete velocity method and the unified gas kinetic scheme, respectively. The dash-dotted and dashed lines are analytical solutions of NSEs~\eqref{per_large_por}, with the no-slip and first-order velocity-slip boundary conditions, respectively. Open circles represent numerical solutions of R20 equations. }
	
	\label{Apparent_Permeability}
\end{figure}

The accuracy of the slip-corrected permeability is assessed by comparing to numerical solutions of the BGK equation and R20 equations, when the porosity is $\epsilon=0.8$. For the linearized BGK equation, two reduced distribution functions were introduced to cast the three-dimensional molecular velocity space into a two-dimensional one, and the obtained two equations are solved numerically by the discrete velocity method~\citep{GraurVacuum2012} and the unified gas kinetic scheme~\citep{Huang2013}. In the unified gas kinetic scheme, a body-fitted structured curvilinear mesh is used, with 150 lines along the radial direction and 300 lines along the circumferential direction, see Fig.~\ref{Apparent_Permeability}(a). In the discrete velocity method, a Cartesian grid with $801\times401$ equally-spaced points is used and the solid surface is approximated by the``stair-case". In solving the R20 equations, a similar body-fitted mesh with $201\times101$ cells is used, and the detailed numerical method is given by~\cite{Gu2009}. The molecular velocity space in the BGK equation is also discretized: $v_1$ and $v_2$ are approximated by the $8\times8$ Gauss-Hermite quadrature when $Kn$ is small ($Kn<0.01$ in this case), and the Newton-Cotes quadrature with $22\times22$ non-uniform discrete velocity points~\citep{lei_Jfm} when $Kn$ is large.

The apparent permeability is plotted in Fig.~\ref{Apparent_Permeability}(b) as a function of $Kn$. When $Kn\lesssim0.15$, our numerical simulations based on the linearized BGK equation and R20 equations agree with each other, and the apparent permeability is  a linear function of $Kn$. When $Kn\gtrsim0.2$, the R20 equations, although being accurate to the third-order of $Kn$, predict lower apparent permeability than that of the BGK equation.  The slip-corrected permeability~\eqref{per_large_por} increases linearly with $Kn$ only when $Kn\lesssim0.02$, and then quickly reaches to a maximum value when $Kn\gtrsim0.2$. This comparison clearly demonstrates that, NSEs with FVBC are only accurate to the first-order of $Kn$. This result is in accordance with the approximation~\eqref{NS_fist} adopted in the derivation of NSEs from the Boltzmann equation. Although the ``curvature of the solid-gas interface'' makes the apparent permeability  a concave function of $Kn$ in the framework of NSEs with FVBC~\citep{Didier2016}, higher-order moments in~\eqref{STRESS}-\eqref{MIJK} and the higher-order velocity slip in~\eqref{ho_velocity}, which are derived from the Boltzmann equation and the gas kinetic boundary condition~\eqref{diffuse} to the third-order accuracy of $Kn$, restore linear dependence of the apparent permeability on $Kn$ when $Kn\lesssim0.15$.

\begin{figure}
	\centering
	\subfloat[]{\includegraphics[scale=0.6,viewport=30 90 555 375,clip=true]{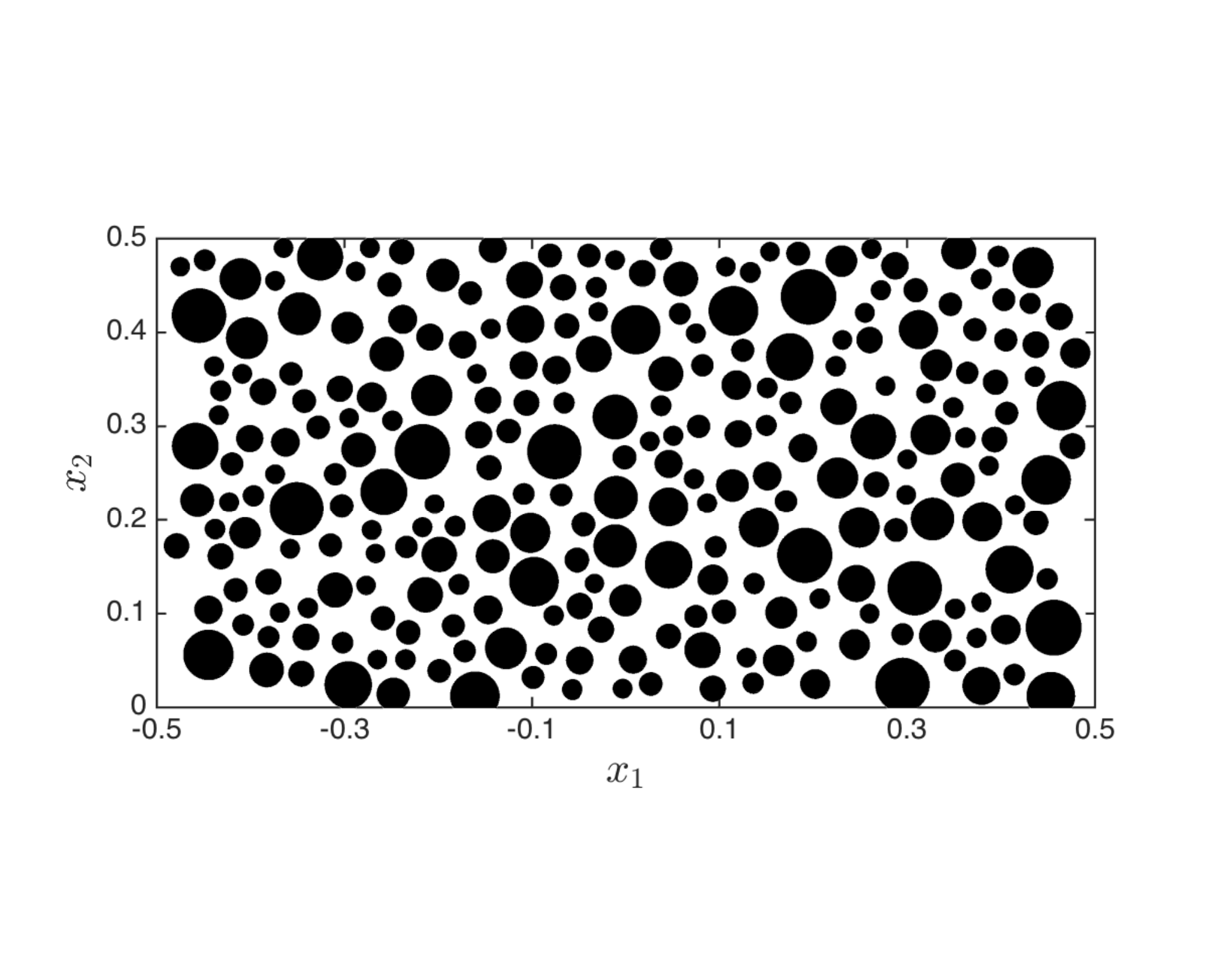}}\\
	
	\subfloat[]	{\includegraphics[scale=0.44,viewport=30 0 620 365,clip=true]{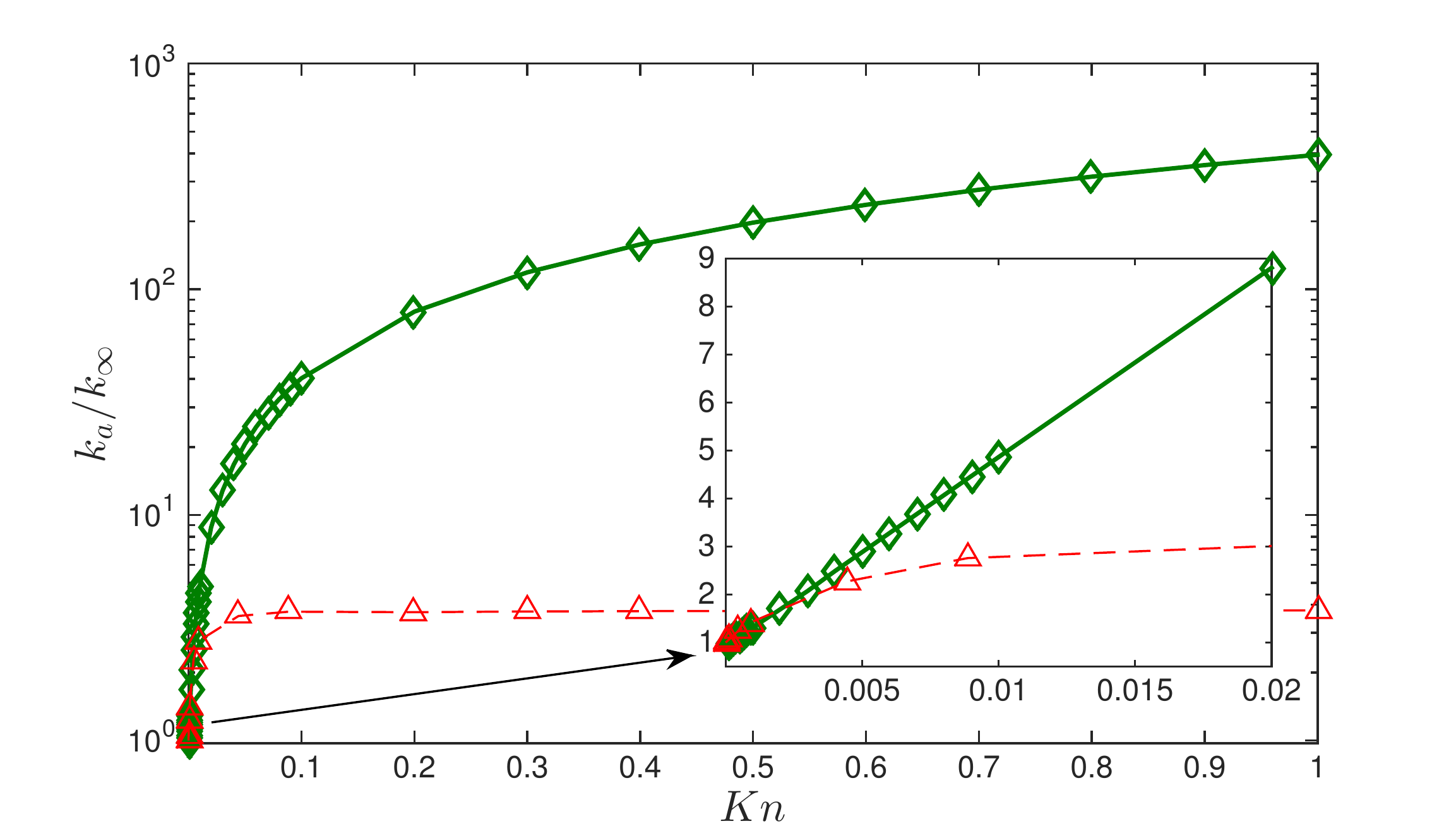}}
	
	\caption{(a) The geometry in a unit computational cell, when the porosity is 0.6. (b) The ratio of the apparent permeability $k_a$ to the intrinsic permeability $k_\infty=9.372\times10^{-6}$ as a function of the Knudsen number. The solid and dashed lines are numerical results of the linearized BGK equation and NSEs with FVBC, respectively.  }	
	\label{RandomDisc}
\end{figure}

The conclusion that NSEs with FVBC is accurate only to the first-order of $Kn$ not only holds for the simple porous medium as shown in Fig.~\ref{GEO}, but also applies to more complex porous media, for example, see the unit cell in Fig.~\ref{RandomDisc}(a) where the porosity is 0.6. In this case, the linearized BGK equation is solved by the discrete velocity method, with a Cartesian mesh of $3000\times1500$ cells; {the grid convergence is verified, as using $6000\times3000$ cells only results in a 0.6\% increase of the apparent permeability when $Kn=1\times10^{-4}$.} NSEs with FVBC are solved in OpenFOAM using the SIMPLE algorithm and a cell-centered finite-volume discretization scheme, on unstructured grids. A body-fitted computational grid is generated using the OpenFOAM meshing tool, resulting in a mesh of  about 600,000 cells of which the majority are hexahedra and the rest few close to the walls are prisms. The apparent permeability from NSEs increases linearly over a very narrow region of the Knudsen number (i.e. $Kn\lesssim0.001$) and then quickly reaches a constant value. Again, from Fig.~\ref{RandomDisc}(b) we see that NSEs with FVBC is roughly accurate when the apparent permeability is a linear function of $Kn$; in this region of $Kn$, the maximum apparent permeability $k_a$ is only about one and a half times larger than the intrinsic permeability $k_\infty$.

\section{Conclusions}

In summary, through our numerical simulations based on the linearized Bhatnagar-Gross-Krook equation and the regularized 20-moment equations, we show that the Navier-Stokes equations with the first-order velocity-slip boundary condition can only predict the apparent permeability of the porous media to the first-order accuracy of the Knudsen number.

\section*{Acknowledgements}
LW acknowledges the support of an Early Career Researcher International Exchange Award from the Glasgow Research Partnership in Engineering, allowing him to visit the Hong Kong University of Science and Technology for one month. LG thanks Lianhua Zhu for helpful discussions on  the first-order velocity-slip boundary condition in OpenFoam. This work is also partly supported by the Engineering and Physical Sciences Research Council in the UK under grant EP/M021475/1.

\appendix

\section{Wall boundary conditions for high-order moments}\label{Wall_hob}

The wall boundary conditions for higher-order moments are given as follows:
\begin{equation*}
\begin{split}
\sigma _{\tau \tau}  =&  - \sqrt {\frac{{\pi RT}}{2}} \left( {\frac{{5{m_{n\tau \tau }} + 2{q_n}}}{{5RT}}} \right) + {p_\alpha }\left( {\hat u_\tau ^2 + {{\hat T}_w} - 1} \right) - \frac{{{R_{\tau \tau }} + {R_{nn}}}}{{14RT}} - \frac{\Delta }{{30RT}} - \frac{{{\phi _{nn\tau \tau }}}}{{2RT}},\\
\sigma _{nn} =&  - \sqrt {\frac{{\pi RT}}{2}} \left( {\frac{{5{m_{nnn}} + 6{q_n}}}{{10RT}}} \right) + {p_\alpha }\left( {{{\hat T}_w} - 1} \right) - \frac{{{R_{nn}}}}{{7RT}} - \frac{\Delta }{{30RT}} - \frac{{{\phi _{nnnn}}}}{{6RT}},\\
q_\tau = & - \frac{5}{{18}}\sqrt {\frac{{\pi RT}}{2}} \left( {7{\sigma _{n\tau }} + \frac{{{R_{n\tau }}}}{{RT}}} \right) - \frac{{5{{\hat u}_\tau }{p_\alpha }\sqrt {RT} \left( {\hat u_\tau ^2 + 6{{\hat T}_w}} \right)}}{{18}} - \frac{{10{m_{nn\tau }}}}{9},\\
m_{\tau \tau \tau } = & - \sqrt {\frac{{\pi RT}}{2}} \left( {3{\sigma _{n\tau }} + \frac{{3{R_{n\tau }}}}{{7RT}} + \frac{{{\phi _{n\tau \tau \tau }}}}{{RT}}} \right) - {p_\alpha }{{\hat u}_\tau }\sqrt {RT} \left( {\hat u_\tau ^2 + 3{{\hat T}_w}} \right) - \frac{{3{m_{nn\tau }}}}{2} - \frac{{9{q_\tau }}}{5},\\
m_{nn\tau }= & - \sqrt {\frac{{\pi RT}}{2}} \left( {{\sigma _{\tau n}} + \frac{{{R_{n\tau }}}}{{7RT}} + \frac{{{\phi _{nnn\tau }}}}{{3RT}}} \right) - \frac{2}{5}{q_\tau } - \frac{{2{{\hat T}_w}{{\hat u}_\tau }{p_\alpha }\sqrt {RT} }}{3},
\end{split}
\end{equation*}
where ${\hat u_\tau } = u_\tau/\sqrt{RT}$ and ${\hat T_w} = T_w/T$.

\bibliographystyle{jfm}

\end{document}